# Cluster expansion made easy with Bayesian compressive sensing

Lance J. Nelson

*Department of Physics and Astronomy, Brigham Young University, Provo, Utah 84602, USA*

Vidvuds Ozoliņš and Fei Zhou

*Department of Materials Science and Engineering,*
*University of California, Los Angeles, California 90095, USA*

C. Shane Reese

*Department of Statistics, Brigham Young University, Provo, Utah 84602, USA*

Gus L. W. Hart

*Department of Physics and Astronomy, Brigham Young University, Provo, Utah 84602, USA*

Long-standing challenges in cluster expansion (CE) construction include choosing how to truncate the expansion and which crystal structures to use for training. Compressive sensing (CS), which is emerging as a powerful tool for model construction in physics, provides a mathematically rigorous framework for addressing these challenges. A recently-developed Bayesian implementation of CS (BCS) provides a parameterless framework, a vast speed up over current CE construction techniques, and error estimates on model coefficients. Here, we demonstrate the use of BCS to build cluster expansion models for several binary alloy systems. The speed of the method and the accuracy of the resulting fits are shown to be far superior than state-of-the-art evolutionary methods for all alloy systems shown. When combined with high throughput first-principles frameworks, the implications of BCS are that hundreds of lattice models can be automatically constructed, paving the way to high throughput thermodynamic modeling of alloys.

## I. INTRODUCTION

Technological advances are driven by the discovery and development of high-performing materials. Discovering these materials is perhaps the single largest bottleneck to technological developments. Due in large part to advances in computing power, computational methods play an increasingly important role in the discovery process. Results from calculations and simulations guide experimental work and provide insight into avenues for future materials research.

The well-known density-functional theory (DFT) is an example of a recent methodological stride in computational materials research. Developed in the 1960's, this theory paved the way to accurate and efficient calculations of materials' properties.[1,2] Steady advances in computing power have made these calculations more affordable computationally, and therefore more viable as a way to probe nature for high-performing materials. This is manifested by recent high-throughput studies that identify new materials and uncover new properties through brute-force calculation of all likely candidates. Results from these studies have been fruitful and illustrative.[3–12]

Although useful for some purposes, high-throughput DFT studies are far from exhaustive in their scope of search and provide no information about the material for $T > 0$. To extend computation's reach, a common approach is to build a much faster model, such as a cluster expansion, and use it to explore $T > 0$ properties via thermodynamic simulations.

Here we employ a "re-weighted Bayesian" implementation of compressive sensing (BCS) to construct cluster expansion models. The CS paradigm addresses, in a mathematically rigorous fashion, two major and long-standing challenges in the cluster expansion community, namely the basis selection problem and the training data selection problem. Although a non-Bayesian implementation of CS[13] provides a solution to these problems using only one adjustable parameter, the reweighted BCS approach removes the adjustable parameter, provides a considerable speed up, and yields sparser models. Most impressive is the fact that re-weighted-BCS-constructed cluster expansion models exhibit a convergence of the solution to a very physical model that predicts more accurately than prevalent methods for the three alloy systems studied in this work.

## II. THE CLUSTER EXPANSION

Cluster expansion provides a fast, accurate way to compute the total energy of all atomic configurations on a parent lattice.[14–16] The cluster expansion is constructed by first assigning each atomic type a pseudo-"spin" variable. Any atomic configuration on the parent lattice can then be specified using a vector of pseudo-spin variables. The physical quantity of interest is then expressed as a linear combination of basis functions, an idea very analogous to a Taylor or Fourier expansion:

$$E(\sigma) = E_0 + \sum_f \bar{\Pi}_f(\sigma) J_f, \qquad (1)$$

where the argument to the function, $\vec{\sigma}$, is a vector of pseudo-spin variables indicating the atomic occupation on the parent lattice sites. The $\bar{\Pi}_f$ are the basis functions, often referred to as cluster functions, with each function corresponding to a cluster of lattice sites. For binary systems, these basis functions are evaluated by averaging over products of pseudo-spin variables. (For higher component systems, the basis is more complex.[14]) The expansion coefficients $J_f$ are called effective cluster interactions (ECI's) and finding their values is the central task when constructing a cluster expansion.

The cluster expansion is essentially a linear algebra problem

$$\bar{\bar{\Pi}}\vec{J} = \vec{E} \qquad (2)$$

with $\vec{E}$ containing the first-principles training data, and $\vec{J}$ the sought-after coefficients, and $\bar{\bar{\Pi}}$ a matrix containing the values of the basis functions evaluated at each training structure. Early in the development of cluster expansion, the ECI's were found by manually truncating the list of admissible cluster coefficients $\vec{J}$ and directly inverting Eq. (2). This so-called structure inversion method (SIM)[17] is conceptually appealing, but in practice the resulting model has poor predictive capability. As the CE method developed, the best practice that emerged was to generate more fitting data than fitting variables (more elements in the vector $\vec{E}$ than in the ECI's vector $\vec{J}$). This results in an overdetermined problem that can be solved, in the least-squares sense, by singular value decomposition or related methods.

Before discussing the fitting approaches in more detail, we point out that whatever the details of the fitting procedure are, any method must deal with two difficulties: (1) The expansion given in Eq. (1) must be truncated to a finite (and typically small) number of terms, and (2) a choice must be made about which structures (among a practically infinite set) should be used as training data (to generate the vector $\vec{E}$). The expansion must be severely truncated so that it has fewer terms than the number of training structures (maintaining an overdetermined problem), and the training structures should be chosen to minimize the predictive errors. Mathematically speaking, the choice of the training structures is not independent of the truncation.

Both difficulties are challenging. The first is difficult because the number of relatively short-ranged clusters is enormous (see Fig. 1) so a robust distance- or hierarchy-based truncation method is not practical. It is difficult to avoid truncating relevant terms inadvertently. There are several contemporary approaches to the truncation problem[18–27]. The second challenge, choosing the structures to be used as training data, is related to the first. The optimal choice of training structures depends on the truncation. Some approaches attempt to choose training structures so as to minimize the variance in predictive errors.[28–30] Others, based on the early work of Garbulsky and Ceder,[31] attempt to bias the training set to repro-

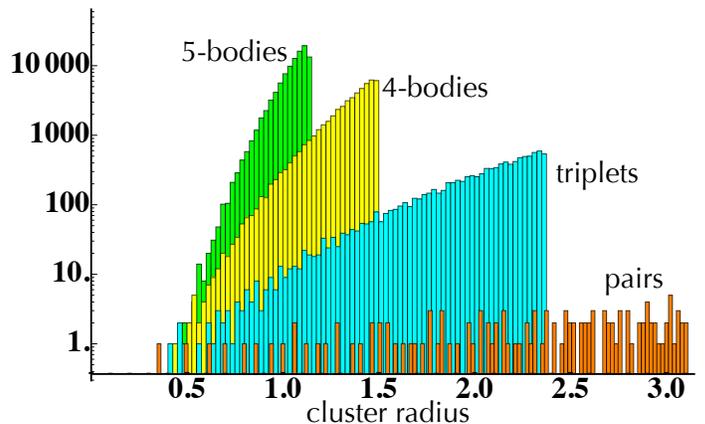

FIG. 1. Partial histogram of geometrically unique clusters on an fcc lattice. The x-axis is the cluster radius, which is defined to be the average distance (in units of lattice constants) from the cluster center of mass to the cluster vertices. The number of unique clusters increases exponentially as the number of cluster vertices and cluster radius increase. This illustrates the magnitude of the challenge associated with truncating the cluster expansion. Note that the histograms have been cut off once the number of clusters becomes very large. This is not meant to imply that there are no clusters beyond this point. Rather the graphic is meant to provide a qualitative view of how quickly the number of unique clusters increases.

duce the correct ordering of low-energy states.[20]

With the exception of recent CE techniques based on Bayesian inference,[27,32,33] the model-building process of contemporary techniques are essentially the same: An initial set of training data is generated and a fit is calculated. The predictive accuracy of the model is assessed. More training data is added and a more refined model is generated. This process is continued, with more and more terms being included in the expansion, until a model with the desired predictive accuracy is achieved.

## III. COMPRESSIVE SENSING

Here we show that compressive sensing (CS), a technique originally developed for applications in signal processing, can be used to select important ECI's and compute their values *in one shot*. To identify important ECIs, CS considers essentially all possible basis functions. Since the number of unique, potentially-relevant clusters is typically very large, considering all possible clusters suggests solving a highly underdetermined version of equation (2) (Many more columns [clusters] than rows [structures] in the matrix $\bar{\bar{\Pi}}$). The compressive sensing cluster expansion (CSCE) method proposed in Ref. 13 solves this heavily under-determined problem by searching for the solution with the smallest $\ell_1$ norm which reproduces the calculated data with a given accuracy,

$$\mathbf{J}_{CS} = \arg\min_{\mathbf{J}}\{\|\vec{J}\|_1 : \|\bar{\bar{\Pi}}\vec{J} - \vec{E}\|_2 < \epsilon\}, \qquad (3)$$

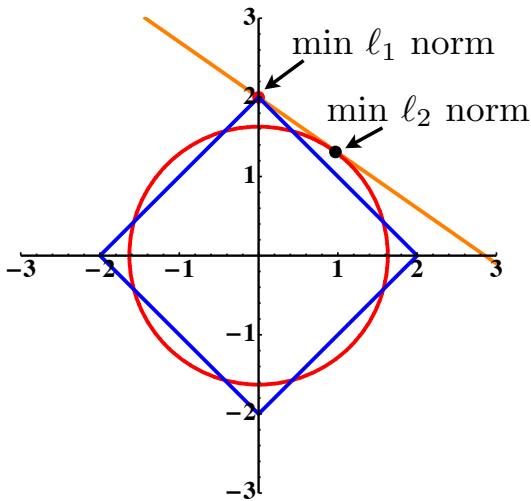

FIG. 2. Illustration of constant $\ell_p$ norm surfaces in $R^2$. The circle is a constant $\ell_2$ norm surface and the diamond is a constant $\ell_1$ norm surface. The straight line indicates the possible solutions to the underdetermined problem $10y + 7x = 20$. A sparse solution to this problem is the solution where one of the variables is zero and the other is not, in other words it is at the intersection of the straight line and the axes. Minimizing the $\ell_2$ norm of this system will result in a dense solution, whereas minimizing the $\ell_1$ norm will yield a sparse solution.

where $\|\vec{J}\|_1$ indicates the $\ell_1$ norm of vector $\vec{J}$, a specific case of the more general $\ell_p$ norm

$$\|\vec{u}\|_p = \left( \sum |u_i|^p \right)^{1/p}. \tag{4}$$

The key idea in compressive sensing is the assumption that the solution vector is sparse, or has few non-zero components. The $\ell_1$ norm constraint, which has been used for years as a sparsity measure, is then used to direct the solution search towards the most sparse solution. Since CE models are known to be sparse, CS provides a fast, robust, and efficient way to detect physically relevant clusters and to compute their corresponding coefficients.[34]

Figure 2 illustrates CS for the simple two-dimensional underdetermined problem $10y + 7x = 20$. The straight line in the figure represents all possible solutions corresponding to this system. The circle (diamond) is a constant $\ell_2$ ($\ell_1$) norm surface. A sparse solution to this system is one where one of the unknowns is non-zero and the other is zero, in other words it is where the straight line intersects one of the axes. The intersection of the solution curve and the constant $\ell_2$ norm curve will always occur off-axis yielding a dense solution. The intersection of the solution curve and the constant $\ell_1$ norm curve will occur on one of the axes, and therefore yield a sparse solution. Constant $\ell_p$ surfaces where $0 < p \leq 1$ can enhance the sparsity, but then finding the global minimum is an NP-hard, non-convex optimization problem.

## A. Training set selection

The mathematical framework of compressive sensing, put forth by Candès, Romberg, and Tao[35], guarantees the recovery of sparse ECI's from a small number of first-principles total energies given certain properties of the matrix $\bar{\bar{\Pi}}$ in Eq. (2). The solution to Eq. (3) was shown to be exact with overwhelming probability if the number of function samples, $m$, satisfies

$$m \geq C \cdot \mu^2(\Phi, \Psi) \cdot S \cdot \log n \tag{5}$$

where $C$ is some positive constant, $n$ is the number of basis functions being considered and $S$ is the sparseness of the solution vector (An $S$-sparse solution vector has $S$ non-zero coefficients). Eq. (5) provides a lower bound on the number of training data points needed to recover the relevant ECIs from a large pool of candidates. The function $\mu(\Phi, \Psi)$ is a measure of the coherence between the sensing basis and the representation basis and is given by

$$\mu(\Phi, \Psi) = \sqrt{n} \max_{1 \leq i, j \leq n} |\langle \phi_k, \psi_j \rangle|. \tag{6}$$

where $\phi_i$ is the representation basis that expresses the signal in a linear model [in our case, the cluster functions $\bar{\Pi}_f$ that represent the energy through Eq. (1)] and $\psi_i$ is the sensing basis used to "sense" or train the linear model. The coherence value $\mu(\Pi, \Psi)$ is bracketed by $[1, \sqrt{n}]$. Ensuring that this function evaluates to its lowest possible value reduces the number of function samples needed to recover the signal and provides a well-defined recipe for choosing training data.

One approach to minimizing coherence is to choose naturally incoherent pairs of bases and then sample the function randomly in the domain of the sensing basis. For example, delta functions and Fourier functions are a maximally incoherent pair, and using the delta functions for the sensing basis and the Fourier functions for the representation basis and sampling the function randomly will result in the function $\mu(\Phi, \Psi)$ being minimal.

In physics applications, the nature of the problem of interest dictates the use of a specific basis. For example, in cluster expansion, the representation basis are the cluster functions and the sensing basis is the specific values of the cluster functions corresponding to the ordered structures in the training set. Where there is no freedom to choose the basis, the best approach is to construct the sensing matrix, in our case $\bar{\bar{\Pi}}$, such that its rows are approximately independent and identically distributed (i.i.d.). This will guarantee that sparse sets of ECI's will not be in the null space of $\bar{\Pi}$, ensuring efficient recovery of the true physical solution. For a more complete description of incoherence as it relates to compressive sensing, see Reference 35.

The simple requirement that the coherence should be minimal provides a mathematically rigorous solution to the question of which structures should be used in the training set. Choosing training structures whose correlation vectors are i.i.d is the best choice when using compressive sensing. This recipe can be applied **once** at the beginning of the model building process instead of using iterative procedures to build up the training set over time. This feature of CS-based CE models provides an automatic and hands-off framework to the model building process.

## B. Training set selection for cluster expansion

For the cluster expansion model, constructing the sensing matrix $\bar{\bar{\Pi}}$ such that its rows are i.i.d is more challenging than in the case of a Fourier expansion. This is because the cluster function values form a discrete, non-uniformly distributed set. Figure 3 gives the distribution of values for the first, second, and third nearest neighbor cluster functions for all fcc-derived superstructures with 12 atoms/cell or less. Clearly the allowed values are not uniformly distributed, and there are values which never occur. Furthermore, the values of the cluster functions are correlated to one another, further complicating the task of choosing training data.

One method for choosing training structures which produces an approximately random sensing matrix was given in reference 13. In that method, vectors of uniformly distributed numbers were first normalized (i.e. random vectors on a hypersphere) and the structure whose vector of cluster functions was closest to this vector was added to the training set.[36]

Another method for accomplishing this involves orthonormalizing the random vectors before matching them to real crystal structures. The exact recipe for doing this proceeds as follows:

### Structure selection procedure

---

1. Generate a random vector $\pi$ on the unit hypersphere.

2. Orthogonalize $\pi$ to all rows of the current sensing matrix $\bar{\bar{\Pi}}$.

3. Normalize $\pi$

4. Find the nearest crystal structure to the orthonormalized $\pi$.

5. Add the structure to the training set.

6. Update the matrix $\bar{\bar{\Pi}}$. Go back to step 1.

---

To investigate which method for picking training data results in the most incoherent, or uncorrelated set of data, several approach were compared: i) picking structures randomly, ii) picking the lowest atom/cell structures in an enumerated list, iii) the approach defined in this work, and iv) the approach of reference 13.

Randomly picked structures were chosen by simply choosing a random integer from 1 to $M$ where $M$ is the number of candidate training structures. The set of lowest atom/cell structures was included to compare to commonly used methods for selecting training data. The quality of each sensing matrix was measured by computing the cross correlation matrix. For $N$ basis functions, we compute an $N \times N$ matrix with elements $\eta_i \cdot \eta_j / (||\eta_i||_2 ||\eta_j||_2)$, where $\eta_i$ is the $i$-th column of the sensing matrix.

The cross-correlation matrix is a simple, but imperfect measure of the ability to discriminate between different pairs of ECIs. Theoretically, a more stringent criterion is the so-called restricted isometry property,[35,37] which guarantees that all $S$-sparse vectors lie outside of the null space of the sensing matrix. However, the latter is difficult to evaluate in practice, and hence we use the much more convenient cross-correlation. We also note that mutual coherence, defined as the maximum absolute value of the off-diagonal elements of the cross-correlation matrix, is used in compressive sensing to characterize the ability to reconstruct the true signal.[38]

For each method 500 training structures were chosen and the associated cross-correlation matrix was constructed. The distribution of the off-diagonal term for each method is shown in figure 4. The distribution for a purely random sensing matrix, essentially optimal for compressive sensing, is also shown in the figure for a reference. Clearly, choosing structure numbers at random leads to the poorest cross-correlation values. Considerable improvement is achieved from using the method of reference 13. Even further improvement is achieved with the method put forth in this paper, and this method will be employed for the current high-throughput work.

## IV. MATHEMATICAL IMPLEMENTATIONS

Various mathematical techniques exist for solving an underdetermined linear system subject to a constraint. One such method recasts the constrained minimization problem of Eq. 3 as the unconstrained minimization problem

$$\min_{\vec{J}} \{\mu \|\vec{J}\|_1 + \frac{1}{2}\|\bar{\bar{\Pi}}\vec{J} - \vec{E}\|_2^2\}. \tag{7}$$

This equation is referred to as the basis pursuit denoising problem, and it can be solved efficiently using algorithms based on the so-called Bregman iteration.[39,40] The sparseness of the solution can be tuned by varying the parameter $\mu$. Smaller (Larger) values of $\mu$ mean that the $\ell_1$-norm term will be weighted less (greater) than the $\ell_2$-norm term and will therefore result in less (more) sparse solutions.

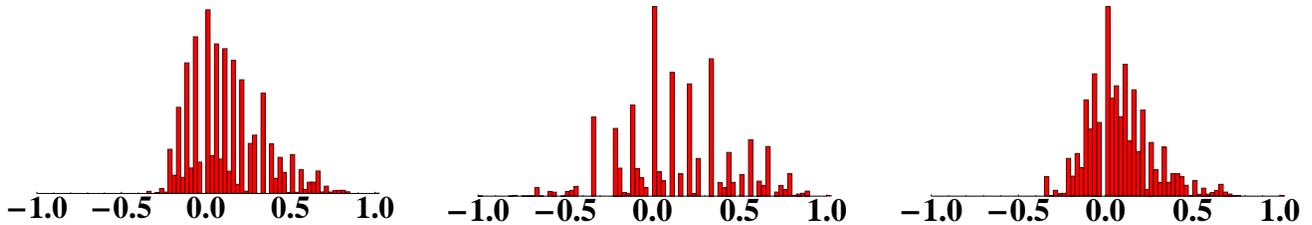

FIG. 3. Histograms of the value of the 1st (left), 2nd (center), and 3rd (right) nearest neighbor pair cluster functions over all fcc-derived superstructures up to 12 atoms/cell. Most noteworthy is the fact that the cluster function values are not uniformly distributed. Also, note that there are regions of values which never occur over this set of structures. These points make it challenging to construct a sensing matrix composed of random, uniformly distributed entries.

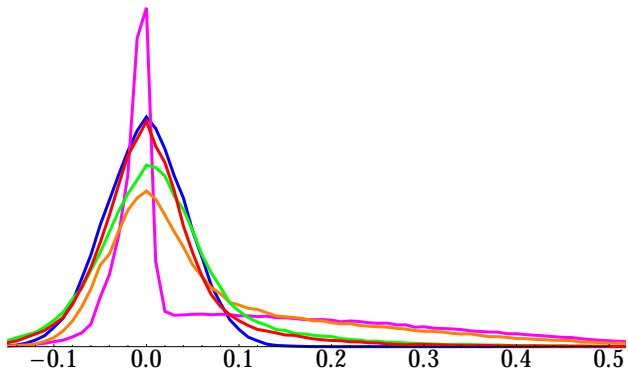

FIG. 4. Distribution of off-diagonal elements of the cross-correlation matrix for several different methods for choosing training data. Choosing structures with the smallest unit cells (magenta), choosing structures at random (orange), choosing structures using the method described in reference 13 (green), and the method discussed in this paper (red) are displayed. The distribution of the cross-correlation off diagonal terms for a purely random matrix is given as a reference (blue).

## A. A Bayesian Implementation

Bayesian inference has been used as a model-building tool[27,32,33] previously, but it's models are typically dense (non-physical) and the framework for building the models can be time intensive for the user. Standard implementations of compressive sensing are powerful tools for model building, but require tuning the sparsity parameter $\mu$.[13] Combining compressive sensing and Bayesian statistics has several significant advantages over traditional Bayesian approaches and the previous CS implementation as well as other prevailing methods:[20,29,41] extreme computational efficiency, sparse models with high predictive accuracy, error estimates for predictions, and the elimination of tunable parameters so that models can be developed automatically, entirely "hands-off."

Here we employ a recently developed Bayesian implementation of CS,[42,43] which leads to a relatively simple numerical algorithm and is based on several key ideas from the CS and Bayesian literature. To illustrate how the numerical algorithm is derived from the starting concepts, we begin first with two short examples to introduce the ideas of conjugacy and sparsity, which play a central role in determining the final form of the "posterior distribution" that the numerical algorithm retrieves.

*a. Bayesian Inference* The foundation of Bayesian inference stems from a simple statement of conditional probability. Consider the joint probability that both events $a$ and $b$ will happen,

$$P(a \cap b) = P(b|a)P(a), \tag{8}$$

but obviously

$$P(b \cap a) = P(a|b)P(b). \tag{9}$$

Equating these two expressions leads to Bayes' theorem

$$P(a|b) = \frac{P(b|a)P(a)}{P(b)}. \tag{10}$$

In words, this theorem states that the probability of $a$ given that $b$ is true is proportional to the probability of $b$ given that $a$ is true. This rule can be easily applied to answer questions involving simple yes/no events. For example, if $a$ corresponds to actually having breast cancer and $b$ corresponds to receiving a positive test result for breast cancer, then the result of Bayes' rule would give the probability that a person who receives a positive test result actually has cancer. In this case, each term in Bayes' rule is a single number, the probability of the corresponding event.

When the problem of interest is not a simple yes/no question, the terms in Bayes' rule become probability distributions (pdf),[44]

$$\mathrm{p}(\mu, \sigma|\vec{x}) \propto \mathrm{p}(\vec{x}|\mu, \sigma)\mathrm{p}(\mu)\mathrm{p}(\sigma). \tag{11}$$

Here, Bayes' rule provides inference on the quantities $\mu$ and $\sigma$ given the information contained in the data, $\vec{x}$. The probability distributions $\mathrm{p}(\mu)$ and $\mathrm{p}(\sigma)$ are called

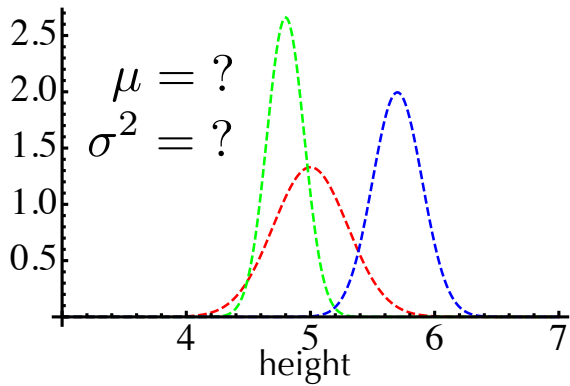

FIG. 5. Illustration of Bayes' rule for heights of college students. One can assume that the distribution is well approximated by Normal (Gaussian) distribution, but the mean and width of the distribution are initially unknown—three possible distributions are shown. Bayes' rule provides distributions on the mean and the width, indicating what values are likely for these parameters.

"prior" distributions and they provide *a priori* estimates on the value of the parameters $\mu$ and $\sigma$. The distribution p($\vec{x}|\mu, \sigma$) is called the likelihood and is the distribution that the data is presumed to have come from.

As a simple application, consider the heights of college students. The distribution of heights is well approximated by a Normal distribution,(i.e. the likelihood is Normal) but the mean and the variance of the distribution are unknown(see Fig. 5). Bayes' rule can be used to estimate the value of these parameters and the first step is to employ a Normal distribution for the likelihood

$$p(\vec{y}|\mu, \sigma^2) = \mathcal{N}(\vec{y}|\mu, \sigma^2). \qquad (12)$$

Since there are two parameters in the likelihood, there must also be two prior distributions, one for each parameter. These distributions, $p(\mu)$ and $p(\sigma^2)$, are *a priori* information about the values of $\mu$ and $\sigma$ and are chosen using physical intuition about the situation. The posterior distribution, $p(\mu, \sigma^2|\vec{y})$, which is formed from the product of the likelihood and prior distributions, appropriately weights prior information and the information provided in the data to provide distributions for the parameters $\mu$ and $\sigma^2$.

*b. Conjugacy* The term conjugacy in Bayesian statistics refers to a specific relationship between the likelihood and the prior. When a prior which is conjugate to the likelihood is chosen, the posterior distribution, which is the product of the likelihood and the prior, belongs to the same family as the prior distribution. Suppose a random sample $y_1, \ldots, y_n$ are collected from a Gaussian distribution. Then computing the product of a Gaussian likelihood

$$\mathcal{N}(\vec{y}|\mu, \sigma^2) = \left(2\pi\sigma^2\right)^{-n/2} e^{-\frac{1}{2\sigma^2}\sum_{i=1}^{n}(y_i-\mu)^2} \qquad (13)$$

and an inverse gamma prior distribution on $\sigma^2$

$$\gamma(\sigma^2|\alpha, \beta) = \frac{\beta^\alpha}{\Gamma(\alpha)}(\sigma^2)^{-\alpha-1}e^{-\frac{\beta}{\sigma^2}} \qquad (14)$$

is formed, the resulting posterior distribution on $\sigma^2$ is also an inverse gamma distribution with the new parameters

$$\alpha_n = \alpha + \frac{n}{2}, \qquad \beta_n = \beta + \frac{\sum_{i=1}^{n}(y_i-\mu)^2}{2}. \qquad (15)$$

Choosing a conjugate prior results in two important advantages over other choices. 1) Computational complexity is reduced and, 2) because the form of the posterior is recognizable, the mean and variance of the posterior are easily identified. By choosing a conjugate prior, the mean and variance of the posterior are known analytically, and if desired, the posterior distribution can easily be sampled. If a non-conjugate prior is chosen, retrieving the posterior distribution requires costly sampling algorithms such as Markov Chain Monte Carlo and Metropolis Hastings.[45–49]

*c. Sparsity* A sparse model is both intuitively appealing and computationally efficient when the model is used in subsequent simulations. The compressive sensing paradigm seeks a *sparse* solution to an underdetermined system by minimizing the $\ell_1$-norm of the solution vector. Most compressive sensing implementations do just that: minimize the $\ell_1$-norm of the solution vector through a convex optimization algorithm with a tunable parameter to adjust the balance between sparsity and the magnitude of the fitting errors. Tunable parameters can be eliminated with BCS but enforcing sparsity is more subtle and relies on a specific choice for the prior distribution on the model coefficients. Combining the strengths of both compressive sensing and Bayesian inference yields a computationally efficient, parameterless algorithm for recovering the most sparse solution. Other Bayesian methods for model recovery are effective,[27,32,33] but the number of parameters in such a model can be large.

### B. A combined Bayesian compressive sensing algorithm

#### 1. A simple example

To illustrate how compressive sensing can be implemented using Bayesian statistics, let's illustrate the technique on the simple underdetermined system of $10y + 7x = 20$, which was briefly mentioned in section III. A logical choice for the likelihood for this problem is a normal centered at $10y + 7x$ with variance $\sigma^2$

$$p(\text{data}|\mu, \sigma^2) = \mathcal{N}(\text{data}|10y + 7x, \sigma^2) \qquad (16)$$

This essentially says that if one knew the values of $x$ and $y$, then one could draw from a normal distribution centered at $10y + 7x$ and having variance $\sigma^2$ and the resulting numbers would approximate the data collected (in our case it is a single data point: 20).

With the likelihood defined, we must now define prior distributions on the parameters found in the likelihood, namely $x$ and $y$.[50] The choice of prior distributions on the parameters $x$ and $y$ is key to implementing the compressive sensing paradigm. One common way to implement the CS paradigm is to use a Laplace distribution for the priors on $x$ and $y$. The Laplace distribution enhances sparsity by placing a large probability mass at the origin, thus favoring 0 for parameter values. However the Laplace distribution is not conjugate to the Normal likelihood and would result in greater computational complexity. To work around this challenge, a Normal distribution is used for the priors on $x$ and $y$,

$$x \sim \mathcal{N}(0, \gamma_1) \qquad y \sim \mathcal{N}(0, \gamma_2) \qquad (17)$$

and Gamma distributions (hyperpriors) are placed on the parameters $\gamma_1$ and $\gamma_2$.[51]

$$\gamma_1 \sim \Gamma(1, \frac{\lambda}{2}) \qquad \gamma_2 \sim \Gamma(1, \frac{\lambda}{2}). \qquad (18)$$

The full bivariate posterior distribution is then assembled as

$$
\begin{aligned}
\mathrm{p}(x, y|\mathrm{data}, \sigma^2) = {} & \mathcal{N}(\mathrm{data}|10x + 7y, \sigma^2)\mathcal{N}(x|0, \gamma_1) \\
& \times \mathcal{N}(y|0, \gamma_2)\Gamma(\gamma_1|1, \frac{\lambda}{2})\Gamma(\gamma_2|1, \frac{\lambda}{2})
\end{aligned} \quad (19)
$$

Notice that when $\lambda \to 0$ the hyperprior is very broad, thus providing very little information about the value of $\gamma_i$ and when $\lambda \to \infty$ the hyperprior becomes a highly-restrictive delta function centered at the origin. If the hyperprior is a delta function, $\gamma_i = 0$ and the prior distribution on coefficient $i$ is also a delta function. This essentially cuts out basis function $i$ from the model. Thus it is easy to see how sparsity can be enforced through this framework.

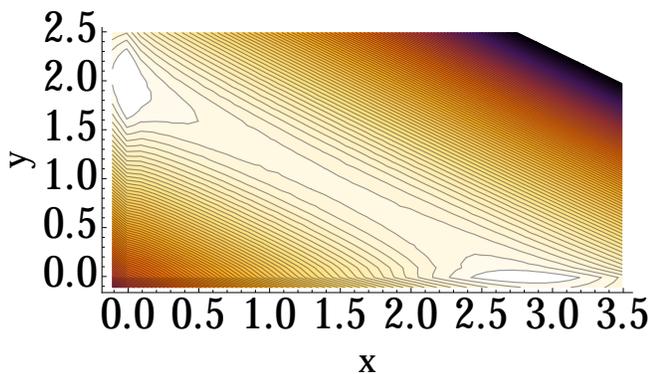

FIG. 6. Contour plot of distribution given in Eq. (19) with $\lambda = 10$ and $\sigma = 3.5$ and $\gamma_1$ and $\gamma_2$ integrated out. The distribution exhibits two peaks: one at $(0, 1.96)$ and the other at $(2.79, 0)$, both sparse solutions to the underdetermined problem $10y + 7x = 20$.

When the $\gamma$'s are integrated out and reasonable values for $\lambda$ and $\sigma^2$ are chosen, the resulting distribution over $x$ and $y$ is shown in Fig. 6.[52] This two dimensional distribution exhibits a peak at the location $(0, 1.96)$, a sparse solution to the underdetermined problem $10y + 7x = 20$. Another peak is also found at the other sparse solution: $(2.79, 0)$, but it is lower in magnitude as it corresponds to a solution with a greater $\ell_1$ norm.

### 2. Algorithmic details

Real problems are of much higher dimensionality than what was illustrated in the previous section. However, the form of Bayes' rule remains essentially unchanged

$$
\begin{aligned}
\mathrm{p}(\vec{J}, \sigma^2, \vec{\gamma}, \lambda|\vec{E}) = {} & \mathcal{N}(\vec{E}|\bar{\Pi}\vec{J}, \sigma^2)\prod_i^N \mathcal{N}(J_i|0, \gamma_i) \\
& \times \left[\prod_i^N \Gamma(\gamma_i|1, \frac{\lambda}{2})\right]\Gamma(\sigma^2|a^\beta, b^\beta)\Gamma(\lambda|\frac{\nu}{2}, \frac{\nu}{2})
\end{aligned} \quad (20)
$$

Here, $\vec{E}$ is a vector of function samples. For cluster expansion models, this would be a vector of first-principles data. The vector $\vec{J}$ contains the sought-after model coefficients. Notice that instead of fixing the values of $\sigma^2$ and $\lambda$ we now introduce prior distributions on these parameters. This introduces several new parameters $a^\beta$, $b^\beta$, and $\nu$, the values of which must be chosen at the outset.

To increase computational speed, the projection of the full posterior distribution in Eq. (20) onto a lower dimensional subspace where we condition on optimal values for nuisance parameters $\sigma^2, \gamma$, and $\lambda$ yields a convenient form for the conditional joint posterior distribution over the model coefficients $\vec{J}$

$$\mathrm{p}(\vec{J}|\sigma^2, \vec{\gamma}, \lambda, \vec{E}) = \mathcal{N}(\vec{J}|\mu, \Sigma). \qquad (21)$$

Here we define the mean vector as

$$\mu = \Sigma\sigma^{-2}\Pi^T\vec{E} \qquad (22)$$

and the covariance matrix as

$$\Sigma = \left[\sigma^{-2}\Pi^T\Pi + \Gamma\right]^{-1}, \qquad (23)$$

where

$$\Gamma = \mathrm{diag}\left(\frac{1}{\gamma_i}\right). \qquad (24)$$

Once accurate values for $\Sigma$ and $\mu$ are known, the resulting distribution provides the sought-after estimate of the ECIs, $\vec{J}$. However, notice that $\Sigma$ and $\mu$ are dependent on the parameters $\gamma$, $\lambda$, and $\sigma^{-2}$. Optimal values for the parameters $\sigma^2, \gamma$, and $\lambda$ can be obtained by first projecting the full posterior of Eq. (20) onto the subspace defined by the variables $\sigma^2, \gamma$, and $\lambda$, thus forming the conditional joint posterior distribution

$$p(\sigma^2, \vec{\gamma}, \lambda | \vec{J}, \vec{E}). \qquad (25)$$

This distribution is then maximized with respect to the variables $\sigma^2, \gamma$, and $\lambda$. What emerges are analytic expressions for $\sigma^2, \gamma$, and $\lambda$, with each expression being dependent on the other two variables. Finding the optimal values for these variables is done using an iterative procedure where the most current version of the set of parameters is used to update the remaining, out-of-date, parameters.

The iterative solution employed here includes the updates of $\Sigma$ (Eq. (23)) and $\mu$ (Eq. (22)) at each iteration. The update of $\Sigma$ would normally require a costly inverse (especially costly for problems involving large cluster pools). To avoid the computationally expensive inverse found in Eq. (23), Babacan *et al.* update a single $\gamma_i$ per iteration. Instead of computing an inverse at each iteration, the entries in the matrix corresponding to the current-iteration basis function are simply updated. This leads to a very efficient update of the matrix $\Sigma$ and the mean vector $\mu$. The speed of this implementation hinges critically on this idea. It is insightful to note that if $\gamma_i = 0$ then $J_i = 0$ and the corresponding model coefficient is 0. Since we expect sparse solutions, many of the $\gamma_i$'s are expected to be zero, and the covariance matrix and mean vector can be represented with far fewer dimensions than $N$.

The algorithm proceeds by beginning with the zero model, all $\gamma_i$'s are set to zero and therefore all model coefficients are zero, and then iteratively adding, removing, or reestimating model coefficients:

---

### Bayesian Compressive Sensing

---

- set all $\gamma_i = 0$

- While not converged do:

  1. Choose a basis function to consider, $\gamma_i$.

  2. Compute the value of $\gamma_i$ which maximizes the posterior distribution (Eq. (25)). Call this value $\gamma_i^{(m)}$.

     - If $\gamma_i^{(m)} < 0$: prune $\gamma_i$ out of the model(set $\gamma_i = 0$).

     - If $\gamma_i^{(m)} > 0$ and $\gamma_i = 0$: Add $\gamma_i$ to the model.

     - If $\gamma_i^{(m)} > 0$ and $\gamma_i > 0$: Re-estimate the value of $\gamma_i$

  3. Update all other parameters.($\Sigma$, $\mu$, $\lambda$, $\sigma^2$)

- end While

---

At step one of the algorithm, a basis function $\gamma_i$ is selected for consideration. This selection is made by computing the value of each $\gamma_i$ and choosing the one that results in the greatest increase in the posterior distribution [Eq. (25)]. The algorithm is stopped when the increase in the posterior distribution from one iteration to the next is less than some predefined threshold.

### C. Enhancing the sparsity through re-weighted $\ell_1$ norm minimization

The $\ell_1$ norm is a practically useful and computationally efficient, albeit less-than-perfect, measure of sparsity. A more accurate measure of sparsity is given by the $\ell_0$ norm, which counts the number of non-zero elements in a vector. However, the $\ell_0$ norm is not a norm in a strict mathematical sense and its use in optimization algorithms is difficult because it is not convex and leads to algorithms that are $NP$-complete.

One drawback with using the $\ell_1$ norm as a measure of sparsity is its dependence on the magnitude of the coefficients. The $\ell_1$ norm favors solutions with smaller-magnitude coefficients over solutions that are equally sparse (or even slightly more sparse), but whose coefficients have larger magnitudes. To address this imbalance, Candes *et al.* proposed a weighted formulation of the $\ell_1$ minimization which penalizes all non-zero coefficients equally.[53] Under this approach the constrained minimization problem is solved iteratively with the $\ell_1$ norm of the model coefficients being weighted at each iteration according to:

$$w_i^{(l+1)} = \frac{1}{|J_i|^{(l)} + \epsilon},\qquad(26)$$

where the index $i$ indicates the basis function being weighted and $l$ is the iteration index. These weights put large and small magnitude coefficients on equal footing by suppressing the contribution of large magnitude coefficients to the $\ell_1$ norm. As explained in reference 53, this weighting can be easily enforced by multiplying the sensing matrix by the inverse of the weight matrix:

$$\bar{\Pi}(W^{(l)})^{-1},\qquad(27)$$

where $W$ is a diagonal matrix with the weights of Eq. (26) on the diagonal. For cluster expansions, we found that re-weighting increases sparsity with a negligible increase in predictive error.

In the absence of the re-weighting procedure, many fits, each using a different training set, must be constructed and the results analyzed statistically to identify dominant coefficients.[13] This increases sparsity and eliminate small, but spurious interactions that result from a particular choice of the training set. However, the re-weighting procedure employed here results in a significant enhancement of sparsity and eliminates the need to average over many solutions.

## V. APPLICATION

Here we demonstrate re-weighted $\ell_1$ minimization through Bayesian compressive sensing on cluster expansion models for the binary systems: Cu-Pt, Ag-Pt, and Ag-Pd. Pt group metal alloys have applications in catalysis and jewelry, which motivated their study here. Additionally, an alternate implementation of CSCE was recently used to study Ag-Pt,[13] and a direct comparison to this alloy was desired.

Using the UNCLE software, approximately 1000 clusters were enumerated, with approximately the same number from each order up to six-body clusters. For each alloy system, the chemical energies of crystal structures were calculated from the density-functional theory (DFT) using the VASP software.[54,55] We used projector-augmented-wave (PAW) potentials[56] and the generalized gradient approximation (GGA) to the exchange-correlation functional proposed by Perdew, Burke and Ernzerhof.[57] To reduce random numerical errors, equivalent $k$-point meshes were used for Brilliuon zone integration.[58] Optimal choices of the unit cells, using a Minkowski reduction algorithm, were adopted to accelerate the convergence of the calculations.[59] The effect of spin-orbit coupling was not included in our calculations because it's effect was shown to be a simple tilt of the calculated energies, as explained in Ref. 60. In total, approximately 800 training structures plus 200 holdout structures with super-cells of up to 12 atoms were calculated for each system

To compare to currently used methods in the cluster expansion community we use the UNCLE code, which uses a genetic algorithm (GA), for the cluster selection/fitting process. GA parameters were set to values that would enable a reasonable computation time and produce typical quality results: 3 populations, 100 generations with 30 children per generation, and a modest mutation rate. While re-weighted BCS is able to consider very large cluster pools, the GA slows considerably as the size of the cluster pool grows. To make a fair comparison, we have used a pool of $\sim 1000$ clusters for both methods. BCS fits for approximately 100 different choices of the training set were performed. Due to the high computation cost of a GA fit, fits for only 5 different training set choices were performed with the GA.

The CS paradigm considers all clusters in the pool equally with no explicit restriction on which, or how many, clusters should be used. To make a fair comparison with the genetic algorithm, the maximum number of model coefficients that the GA was allowed to use was set to be 500. In every fit depicted here, the number of model coefficients found was less than 500.

Since the predictive errors are Gaussian-distributed with a mean of zero, the statistical uncertainty in the predictive error due to the finite size of the prediction set (between 150 and 600 holdout data points) can be calculated using standard statistical formulas for the $\chi^2$-distribution; they are found to be less than 5% of the calculated RMSE.

Figure 7 gives a comparison between GA fits and re-weighted BCS fits for the binary systems Cu-Pt, Ag-Pt, and Ag-Pd. Notice that for every system the root-mean-square error(rmse) over the holdout set is lower for BCS fits for all sizes of the training set. While the rmse of the GA fits is not terrible, the $\ell_1$-norms of the solution vector for GA fits are considerably larger than those from BCS-fits. Furthermore, the $\ell_0$ norm of the GA solutions increases steadily with the size of the training set while the $\ell_0$ norm of the BCS fits remains flat. Clearly, the GA solutions are much more dense than what BCS finds. A dense model is not consistent with widely-held intuition about the nature of physical interactions in real solids. Furthermore, using a many-parameter model for subsequent thermodynamic simulations will result in unnecessary computational burden and prolonged simulation times. In contrast, the $\ell_1$-norm for BCS fits is relatively small and levels off as more training data is added. This is convincing evidence that the solution is converging, and the physical model is being recovered. A graphical comparison of BCS with the CS implementation of Reference 13 is not given because the predictive capacity and $\ell_1$ norm of the solutions are very similar. The main differences are the removal of a tunable parameter and the addition of the reweighting procedure which dramatically reduces the $\ell_0$ norm of the solution

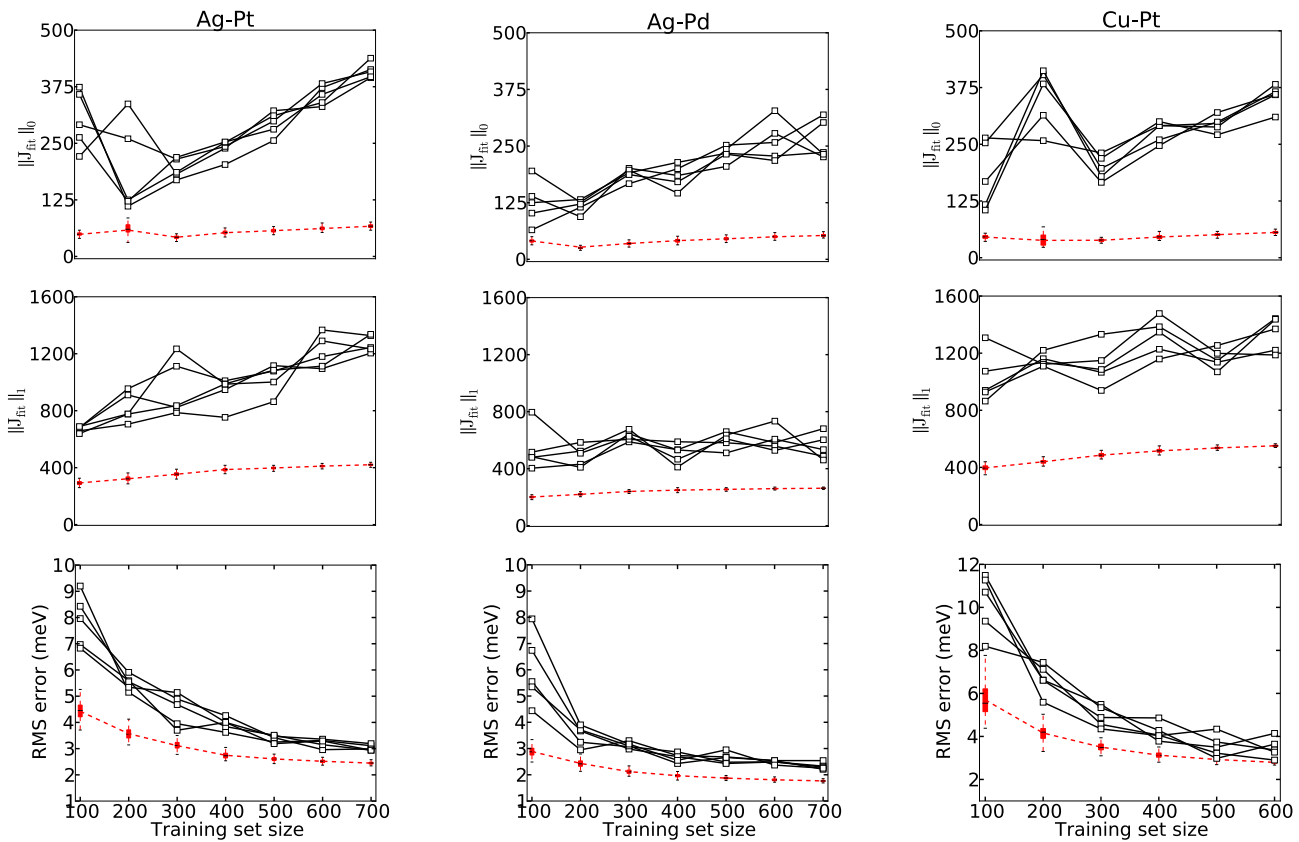

FIG. 7. Comparison between re-weighted Bayesian compressive sensing and genetic algorithm methods for constructing a cluster expansion model for the binary systems Ag-Pt, (left) Ag-Pd, (center) and Cu-Pt (right). The dashed curves indicate BCS results and the solid curves indicate GA results. The upper plot show the $\ell_0$ norm of the solution vector as the training set increases. The middle plot show the $\ell_1$ norm of the solution vector, and the lower plot gives the rmse over a holdout dataset. Approximately 100 BCS fits were performed at each training set size, and the results of these fits are depicted using box-and-whiskers. Due to it's high computational cost, only 5 GA fits were performed, and hence GA results are not depicted using box-and-whiskers.

It is curious that the BCS and GA models achieve similar predictive capacities but differ wildly in the nature of their solutions. One possible explanation for this is that since the GA does not limit the $\ell_1$ norm of the solution vector, its solutions are dense and encompass an approximate null space. Hence, approximate linear dependencies will exist between ECIs of a dense solution, but are much less likely for sparse solutions, like those found by compressive sensing. This could explain how contributions from large ECI coefficients may cancel each other and result in relatively small RMS errors, but this issue certainly needs to be investigated further.

Another key feature of BCS is the efficiency of the algorithm. For the three systems discussed here BCS fits were constructed in a fraction of the time needed for the GA. BCS required on the order of minutes to construct 100 fits, whereas the GA needed $\sim 24$ hours for a single fit.

## VI. CONCLUSION

It has been shown that the CS paradigm is uniquely well-suited to building CE lattice models. Re-weighted BCS-based provides a fast, efficient, and parameterless framework for constructing CE models. These models are constructed in a fraction of the time required by current state-of-the art techniques and with minimal time and effort required by the user. BCS-constructed CE models converge to solutions which agree with widely-held intuition about the nature of physically relevant interactions and predict more accurately than other modern CE construction methods.

From a broader perspective, the CS paradigm is poised to have a big impact on computational physics problems of all types. The CS-paradigm is well suited to tackle any highly-underdetermined linear problem: $\mathbb{A}\vec{x} = \vec{b}$ where $\vec{x}$ is known to be sparse. One possible application is the expansion of high-throughput databases to include lattice models. This approach relies heavily on being able to automatically perform first-principles calculations, and has hitherto not involved using the database information to build materials models. This is mostly due to the high human time cost required to construct such models. However, the hands-off nature of BCS-based CE models will allow materials models to be added to the high-throughput scope of work. In addition to vast amounts of first-principles data, soon high-throughput databases will include accurate lattice models for a diverse array of materials.

## ACKNOWLEDGMENTS

G. L. W. H. and L. J. N. are grateful for financial support from the NSF, DMR-0908753. C. S. R. is grateful for financial support from the NSF, ATM-0934490. F.Z. and V.O. gratefully acknowledge financial support from the NSF under Award Number DMR-1106024 and use of computing resources at the National Energy Research Scientific Computing Center, which is supported by the US DOE under Contract No. DE-AC02-05CH11231.